\documentclass[12pt]{iopart}
\usepackage{latexsym}

\def\tr{{\rm tr}}
\def\ket#1{\mid~\!\!\!{#1}~\!\!\rangle}
\def\bra#1{\langle~\!\!{#1}~\!\!\!\mid}
\def\average#1{\langle~\!\!{#1}~\!\!\rangle}

\begin{document}\jl{1}

\title[Coherence information]{A quantum measure
of coherence and incompatibility}

\author{F Herbut\footnote[1]{E-mail:
fedorh@infosky.net}}
\address{Serbian Academy of
Sciences and Arts, Knez Mihajlova 35,
11000 Belgrade, Serbia and Montenegro}

\date{\today}

\begin{abstract}
The well-known two-slit interference is
understood as a special relation between
observable (localization at the slits) and state
(being on both slits). Relation between an
observable and a quantum state is investigated in
the general case. It is assumed that the amount
of coherence equals that of incompatibility
between observable and state. On ground of this,
an argument is presented that leads to a natural
quantum measure of coherence, called "coherence
or incompatibility information". Its properties
are studied in detail making use of 'the mixing
property of relative entropy' derived in this
article. A precise relation between the measure
of coherence of an observable and that of its
coarsening is obtained and discussed from the
intuitive point of view. Convexity of the measure
is proved, and thus the fact that it is an
information entity is established. A few more
detailed properties of coherence information are
derived with a view to investigate final-state
entanglement in general repeatable measurement,
and, more importantly, general bipartite
entanglement in follow ups of this study.
\end{abstract}


\maketitle

\normalsize

\rm

\section{Introduction}

In a preceding article \cite{FHPR02} {\it
coherence in a relative sense}, i. e., understood
as a relation between a given observable and a
given quantum state, was postulated to be {\it
identical with incompatibility} between
observable and state as far as its {\it quantity}
$I_C$ is concerned. (For notation see the passage
immediately following the proof of Proposition 5
below.) Then it was shown that bipartite pure
state entanglement is expressible as $I_C$ (with
a suitable observable).

Pure states cannot be obtained as mixtures.
Therefore, the question if $I_C$ is concave, i.
e., a genuine entropy quantity, or convex, i. e.,
a genuine information one, or something third,
could not be put in this context. The first aim
of this study is to clarify this point. (This is
done in Proposition 5.) To enable this, the
mixing property of relative entropy (paralleling
the mixing property of entropy and Donald's
identity for relative entropy, see the Remark) is
derived.

In a follow up of the mentioned article
\cite{ent-meas} the special case of the final
bipartite pure state $\ket{\psi}_{12}$ in
repeatable measurement, when the initial state is
pure, was studied. It was shown that the initial
quantity of incompatibility between the measured
observable and the initial state reappears as the
amount of entanglement in $\ket{\psi}_{12}$, and
is further preserved when it is shifted in
reading the measurement result. This completes
Vedral's result \cite{Vedral} that the
information transfer from object (subsystem $1$)
to measuring apparatus (subsystem $2$) does not
exhaust the mutual information $I_{12}$ in the
final state.

I think it is of interest to find out if the
mentioned preservation of the quantity of
incompatibility between the measured observable
and the initial pure state is restricted to pure
state, or it can be generalized to mixed initial
state. This is not a straightforward
generalization. It requires more knowledge on
$I_C$. The second aim of this study is to provide
such knowledge, which will be possible due to the
mentioned auxiliary relative-entropy relations
(see section 3).

In a further preceding article \cite{Roleof} an
arbitrary discrete incomplete observable $A$ and
its completion $A^c$ to a complete observable
were investigated and it was shown that
$I_C(A,\rho ) \leq I_C(A^c,\rho )$ for any state
$\rho$. This inequality is expected if the
assumption on the identity of the amount of
coherence and that of incompatibility is correct.
But it is desirable to evaluate $I_C(A^c,\rho
)-I_C(A,\rho )$ and thus to try to acquire more
insight into the nature of $I_C$. This is the
third aim of this article. (See the discussion
after the proof of the theorem below.)

The fourth aim of this paper is to present an
argument that starts with the mentioned identity
assumption and leads to an expression for the
quantity of coherence in a natural way. Will this
expression be the same as the {\it ad hoc}
introduced one? This is done in section 2 and an
affirmative answer is obtained. It is summed up
in the conclusion (subsection 5.2.).

The fifth and last aim of this investigation is
perhaps the most important one. Namely, in
\cite{Roleof} it was established that $I_C$ plays
an important role also in some mixed bipartite
states. This line of research should be continued
in a follow up because it may contribute to our
understanding how mutual information in general
bipartite states breaks up into a quasi-classical
part and entanglement, which is the object of
study of a wide circle of researchers, e. g.
\cite{VedralHend}, \cite{ZurekOliv}. To this
purpose, one may need more detailed knowledge of
the properties of $I_C$. To acquire such
knowledge is the fifth aim of this article (see
section 4).

\subsection{Background in classical
statistical physics}

To obtain a background for our quantum
study of coherence, we assume that a
classical discrete variable $\quad
A(q)=\sum_la_l\chi_l(q)\quad$ is given
(all $a_l\in {\bf R}$ being distinct).
The symbol $q$ denotes the continuous
state variables (as a rule, it consists
of twice as many variables as there are
degrees of freedom in the system);
$\chi_l$ are the characteristic functions
$\quad\forall l:\quad \chi_l(q)\equiv
1\quad$ if $\quad q\in {\cal A}_l\quad$,
and zero otherwise. Naturally, ${\cal
A}_l$ are (Lebesgue measurable) sets such
that $A(q)=a_l$ if and only if $q\in
{\cal A}_l$, and $\quad \sum_l{\cal
A}_l={\cal Q},\quad$ where ${\cal Q}$ is
the entire state space (or phase space)
and the sum is the union of disjoint
sets.

Let $\rho(q)$ be a continuous probability
distribution in $\cal Q$ with the
physical meaning of a statistical 'state'
of the system. One can think of $\rho(q)$
as of a mixture
$$\rho(q)=\sum_lp_l\rho_l(q),\eqno{(1)}$$
where $\quad \forall l:\quad p_l\equiv
\int_{\cal Q}\rho(q)\chi_l(q)dq\quad$ are
the statistical weights (probabilities of
the results $a_l$ if $A(q)$ is measured
in $\rho(q)$), and $\quad \forall
l,\enskip p_l>0:\quad \rho_l(q)\equiv
\rho(q)\chi_l(q)/p_l\quad$ are the
'states' with definite (or sharp) values
of $A(q)$.

Let $B(q)$ be any other continuous or
discrete variable. Then, utilizing (1),
its average can be written
$$\average{B}_{\rho}\equiv \int_{\cal Q}\rho(q)
B(q)dq=\sum_lp_l
\average{B}_{\rho_l}.\eqno{(2)}$$ {\it
One distinguishes the contributions of
the individual eigenvalues $a_l$ of
$A(q)$ through the terms on the RHS.}
They contribute to $\average{B}_{\rho }$
each separately.

All this serves only as a classical
background to help us to understand the
non-classical, i. e., purely quantum
relations between the analogous quantum
entities.

\subsection{Transition to the quantum mechanical case}

The quantum mechanical analogues of the
mentioned classical entities are the
following.

 Discrete observables
(Hermitian operators) $A=\sum_la_lP_l$
(spectral form in terms of distinct
eigenvalues), $\rho$ quantum state
(density operator), and $B$ an arbitrary
observable (Hermitian operator). The
quantum average is $\quad
\average{B}_{\rho}\equiv \tr (\rho B)$.

In the transition from classical to
quantum one runs into a surprise, that is
known but, perhaps, not sufficiently well
known. Before we formulate it in the form
of a lemma, let us introduce the
L\"{u}ders state $\rho_L$ \cite{Lud} in
order to obtain the quantum analogues of
relations (1) and (2). It is that mixture
of states, each with a definite value of
$A$, which has a {\it minimal}
Hilbert-Schmidt distance from the given
state $\rho$ \cite{rhorho'}. It is
defined as
$$\rho_L\equiv \sum_lp_l\rho_L^l,
\eqno{(3a)}$$ where
$$\forall l:\quad p_l\equiv \tr (\rho
P_l)\eqno{(3b)}$$ are again the
statistical weights in (3a) (or the
probabilities of the results $a_l$ when
$A$ is measured in $\rho$), and
$$\forall l,\enskip p_l>0:\quad
\rho^l_L\equiv P_l\rho
P_l/p_l\eqno{(3c)}$$ are the states with
definite values $a_l$ of $A$. Finally,
$$\average{B}
_{\rho_L}=\sum_lp_l\average{B}_{\rho^l_L}
. \eqno{(3d)}$$

Decomposition (3a) is the analogue of
(1), and (3d) is that of (2).

{\bf Lemma 1.} {\it The following four
statements are equivalent:

(i) The state $\rho$ cannot be written as
a mixture of states in each of which the
observable $A$ has a definite value.

(ii) The observable $A$ and the state
$\rho$ are incompatible, i. e., the
operators do not commute $[A,\rho]\not=
0$.

(iii) The L\"{u}ders state $\rho_L$ given
by (3a)-(3c) is distinct from the
original state $\rho$.

(iv) There exists an observable $B$ such
that
$$\average{B}_{\rho }\not=
\average{B}_{\rho_L},\eqno{(4)}$$ where
the RHS is given by (3d).}

Proof is given in Appendix 1.

The physical meaning of lemma 1 is that  it
defines a kind of {\it quantum coherence} as a
special relation between observable and state.
Experimentally it is exhibited in {\it
interference}. In this relative sense (relation
between variable and state) it is lacking in
classical physics because there a state can
always be written as a mixture of states in each
of which the variable in question has a definite
value (negation of (i), cf (1)). Though classical
waves do exhibit a kind of coherence and show
interference, but this is in a different sense
(cf section 5).

One should note that the L\"{u}ders state needs
no other characterization than its role in lemma
1 (in particular (iii)). The fact that it is
"closest" to $\rho$ in Hilbert-Schmidt metrics,
though actually not important for this study,
raises the thought-provoking questions if
"closest" is true also in other metrics; if not,
why is the Hilbert-Schmidt metrics more suitable.

We take {\it two-slit interference} \cite{Young}
to serve as an illustration for lemma 1.

Let $A$ be a dichotomic position observable with
two eigenvalues: localization at the left slit,
and localization at the right slit on the first
screen. Let $\rho$ be a wave packet that has just
arrived at this two-slit screen. Next, one has to
find a suitable observable $B$ such that
inequality (4) be satisfied at the mentioned
moment. Moreover, one wants to observe
experimentally the LHS of (4), or rather the
individual probabilities of the eigenvalues of
$B$ (that go into the LHS).

To this purpose, one actually replaces $B$ by
another localization observable $A'$ on a second
screen, to which the photon will arrive some time
later. This observable is suitable for
observation (of its localization probabilities).
Hence, one can define $\quad B\equiv
U^{-1}A'U,\quad$ $U$ being the evolution operator
expressing the movement of the particle from the
two-slit screen to the second one. One should
note that $B$ is not a position observable though
$A'$ is because the hamiltonian that generates
$U$ contains the kinetic energy (square of linear
momentum).

Claim (i) of lemma 1 says that the
particle is not moving through either the
left or the right slit. Claim (ii)
expresses the same fact algebraicly.
Namely, $\rho$, being a pure state
$\ket{\psi }\bra{\psi }$, would commute
with $A$ only if $\ket{\psi }$ lay in an
eigensubspace of $A$. In our case this
would mean that the particle traverses
one of the slits.

The L\"{u}ders state $\rho_L$ is, in some
sense, the best approximation to $\rho$
of a state traversing one or the other of
the slits. Naturally, $\quad \rho \not=
\rho_L\quad$ as claimed by (iii). Claim
(iv), i. e., relation (4), amounts to the
same as the fact that the interference
pattern on the second screen is not equal
to the sum of those that would be
obtained when only one of the slits were
open (for some time) and then the other
(for another, disjoint, equally long
time).

In the two-slit experiment one actually
observes the time-delayed equivalent of
(4):
$$\average{A'}_{U\rho U^{-1}}\not=
\average{A'}_{U\rho_LU^{-1}}.\eqno{(5)}$$ Since
the LHS of (5) is {\it distinct} from the RHS,
one speaks of the former as {\it interference}.
In the described two-slit case the LHS of (5)
gives fringes, whereas the RHS does not.
Nevertheless, it is not always true that the LHS
of (5) itself means interference. This is the
case only with a suitable pair of $A$ and $\rho$
(cf (ii) in lemma 1). Let me give a
counterexample.

Let us take another two-slit experiment in which
the slits have polarizers that give opposite
linear polarization to the light passing the
slits \cite{HZ}. The state $\rho$ in the slits is
then such that we have equality in (5) (though
$A'$ is the same), and there is no interference
because $[A,\rho ]=0$. (The state $\rho
=\ket{\psi }\bra{\psi }$ is now in the composite
spatial-polarization state space, and the spatial
subsystem state - the reduced statistical
operator - is a L\"{u}ders state.)

One should note that when interference is
displayed, one has three ingredients: the state
$\rho$, the observable $A$ the two eigenvalues of
which {\it play a cooperative role}, and the
second observable $A'$ the probabilities of
eigenvalues of which are observed. Since in
theory there can be many observables like $A'$,
or $B$ in (4), one likes to omit them. Then one
speaks of {\it coherence} of the observable $A$
in the state $\rho$. We make use of the same
concepts in the general theory.

{\bf Definition 1.} {\it The LHS of relation (4),
in case inequality (4) is valid, is called
interference. If an observable $A$ and a state
$\rho$ stand in such a mutual relation that any
of the four claims of lemma 1 is known to be
valid, then one speaks of coherence.}

One should note that the concepts of interference
and of coherence stand in a peculiar relation to
each other: There is no coherence (between $A$
and $\rho$) unless an observable $B$ that
exhibits interference can be, in principle,
found; if the latter is the case, and only then,
one may forget about $B$, and concentrate on the
relation between $A$ and $\rho$, i. e., on
coherence. The kind of quantum coherence
investigated in this paper can be more fully
called "eigenvalue coherence of an observable in
relation to a state" in view of the cooperative
role of some eigenvalues (or, more precisely,
their quantum numbers, because the values of the
eigenvalues play no role) as seen in (4).

Thus, any of the four (equivalent) claims in
lemma 1 defines coherence. But for the
investigation in this article the important claim
is (ii): coherence exists if and only if $A$ and
$\rho$ do not commute. This remark is the corner
stone of the expounded approach to investigating
coherence (as in the preceding studies
\cite{FHPR02}, \cite{Roleof}).

\section{How to obtain a quantum measure
of coherence?}

We start with the assumption that
coherence of an observable $A$ with
respect to a state $\rho$ is {\it
essentially the same thing} as
incompatibility of $A$ and $\rho$:
$[A,\rho ]\not= 0$. The quantum measure
will be called {\it coherence} or
incompatibility {\it information}, and it
will be denoted by $I_C(A,\rho )$ or
shortly $I_C$ (cf (10) below).

One wonders what the meaning of a larger value of
$I_C$ for coherence is. It is more of what? The
only answer I can think of is in accordance with
the above assumption: More of incompatibility of
$A$ and $\rho$.

The next question is: Do we know what is
a "larger amount of incompatibility"?

The seminal review on entropy of Wehrl
\cite{Wehrl} (section III.C there)
explains that each member of the
Wigner-Yanase-Dyson family of skew
informations $$I_p(\rho ,A)\equiv
-S_p(\rho ,A)\equiv (1/2)\tr ([\rho^p,A]
[\rho^{1-p},A]),\qquad 0<p<1,\eqno{(6)}
$$ is a good measure of incompatibility
of $\rho$ and $A$. Namely, $I_p(\rho ,A)$
is positive unless $\rho$ and $A$
commute, when it is zero. It is also
convex as an information quantity should
be.

Substituting the spectral form of $A$ in
(6), one obtains $$I_p=(1/2)\tr (\sum_l
\sum_{l'}a_l[\rho^p,P_l]a_{l'}[\rho^{1-p},P_{l'}]).
$$ One can see that $I_p$ depends on the
eigenvalues of $A$.

As well known, $A$ and $\rho$ are
compatible if and only if all
eigenprojectors $P_l$ of the former are
compatible with the latter. The
eigenvalues of $A$ do not enter this
relation. Hence, $I_p(\rho ,A)$ given by
(6) is not the kind of incompatibility
measure that we are looking for. One
wonders if there is any other kind.

To obtain an answer, we turn to a
neighboring quantity: the quantum amount
of {\it uncertainty} of $A$ in $\rho$. It
is the entropy $S(A,\rho )$: $$S(A,\rho
)\equiv H(p_l),\eqno{(7a)}$$ where
$H(p_l)$ is the Shannon entropy
$$H(p_l)\equiv
-\sum_lp_llogp_l,\eqno{(7b)}$$ and
$$\forall l:\quad p_l\equiv \tr (P_l\rho
).\eqno{(7c)}$$.

It is known that whenever $A$ and $\rho$
are incompatible, and $A$ is a complete
observable, i. e., if all its eigenvalues
are nondegenerate (we'll write it as
$A^c$), then always $S(A^c,\rho )>S(\rho
)$. When $A^c$ is compatible with $\rho$,
the two quantities are equal. The
interpretation that the larger the
difference $S(A^c,\rho )-S(\rho )$, the
more incompatible $A^c$ and $\rho$ are
seems plausible. Hence, we require for
complete observables $A^c$, that
$I_C(A^c,\rho )$ should equal this
quantity: $I_C(A^c,\rho )\equiv
S(A^c,\rho )-S(\rho )$. Equivalently, one
can require that the following peculiar
decomposition of the entropy in case of a
complete observable should hold:
$$S(\rho )=S(A^c,\rho )-I_C(A^c,\rho
).\eqno{(8)}$$

On the other hand, if $A$ is a discrete
observable that is complete or incomplete
but {\it compatible} with $\rho$, then
the following decomposition parallels
(8):
$$S(\rho )=S(A,\rho )+ \sum_lp_lS(P_l\rho
P_l/p_l)\eqno{(9)}$$ (cf (7a), (7b) and
(7c)). If  $p_l=0$, the corresponding
term in the sum is by definition zero.

Decomposition (9) is obtained by application of
{\it the mixing property of entropy} \cite{Wehrl}
(see Sections II.F. and II.B. there). It applies
to {\it orthogonal state decomposition}, in this
case to $\quad \rho =\sum_lp_l(P_l\rho
P_l/p_l),\quad$ and it reads $\quad S(\rho
)=H(p_l)+\sum_lp_lS(P_l\rho P_l/p_l)\quad$ (cf
(7b)).

The coherence information $I_C$ does not
appear in (9). This is as it should be
because it is zero due to the assumed
compatibility of $A$ and $\rho$.

In case of a general discrete $A$, which
is complete or incomplete, compatible
with $\rho$ or not, we must interpolate
between (8) and (9). This can be done by
observing that both decompositions can be
rewritten in a unified way as
$$I_C(A,\rho )=S\Big(\sum_lP_l\rho
P_l\Big)-S(\rho )\eqno{(10)}$$ (valid for either
$A=A^c$ or for $[A,\rho ]=0$). The searched for
interpolated formula should thus be the same
relation (10), but valid this time for all
discrete $A$. Thus, $I_C(A,\rho )$ is obtained by
the presented argument.

Making use of the mixing property of
entropy, we can rewrite (10) equivalently
as the following general decomposition of
entropy:
$$S(\rho )=S(A,\rho )+ \sum_lp_lS(P_l\rho
P_l/p_l)-I_C(A,\rho ). \eqno{(11)}$$
(Note that $A$ is any discrete observable
in (11).)

In order to derive a number of properties
of coherence information, we make a
deviation into relative entropy theory.

\section{Useful relative-entropy
relations}

The {\it relative entropy}
$S(\rho||\sigma)$ of a state (density
operator) $\rho$ with respect to a state
$\sigma$ is by definition
$$S(\rho||\sigma)\equiv \tr [\rho log(\rho )]-\tr
[\rho log(\sigma)]\eqno{(12a)}$$
$$\mbox{if}\quad \mbox{supp}(\rho ) \subseteq
\mbox{supp}(\sigma );\eqno{(12b)}$$ or
else $\quad S(\rho||\sigma)=+\infty
\quad$ (see p. 16 in \cite{O-P}). By
'support', denoted by 'supp', is meant
the subspace that is the topological
closure of the range.

If $\sigma$ is singular and condition
(12b) is valid, then the orthocomplement
of the support (i. e., the null space) of
$\rho$, contains the null space of
$\sigma$, and both operators reduce in
supp$(\sigma )$. Relation (12b) is valid
in this subspace. Both density operators
reduce also in the null space of
$\sigma$. Here the $log$ is not defined,
but it comes after zero, and it is
generally understood that zero times an
undefined quantity is zero. We'll refer
to this as {\it the zero convention}.

The more familiar concept of (von
Neumann) quantum entropy, $S(\rho )\equiv
-\tr [\rho log(\rho )]$, also requires
the zero convention. If the state space
is infinite dimensional, then, in a
sense, entropy is almost always infinite
(cf p.241 in \cite{Wehrl}). In
finite-dimensional spaces, entropy is
always finite.

There is an {\it equality for entropy}
that is much used, and we have utilized
it, {\it the mixing property} concerning
{\it orthogonal state decomposition} (cf
p. 242 in \cite{Wehrl}):

$$\sigma =\sum_k w_k\sigma_k,\eqno{(13)}$$
$\forall k:\enskip w_k\geq 0$; for $w_k>0$,
$\sigma_k>0,\enskip \tr \sigma_k=1$; $\forall
k\not= k': \sigma_k\sigma_{k'}=0$; $\sum_kw_k=1$.
Then $\quad S(\sigma )=H(w_k)+
\sum_kw_kS(\sigma_k),\quad$ $H(w_k)\equiv
-\sum_k[w_klog(w_k)]\quad$ being the Shannon
entropy of the probability distribution
$\{w_k:\forall k\}$.

The first aim of this section is to
derive an analogue of the mixing property
of entropy. The second aim is to derive
two corollaries that we shall need in
this paper.

We will find it convenient to make use of
an {\it extension} $log^e$ of the
logarithmic function to the entire real
axis: $\quad \mbox{if}\quad 0<x:\qquad
log^e(x)\equiv log(x)\quad$, $\quad
\mbox{if}\quad x\leq 0:\enskip
log^e(x)\equiv 0\quad$.

The following elementary property of the
extended logarithm will be utilized.

{\bf Lemma 2.} {\it If an orthogonal
state decomposition (13) is given, then
$$log^e(\sigma )
=\sum'_k [log(w_k)]Q_k+\sum'_k log^e
(\sigma_k),\eqno{(14)}$$ where $Q_k$ is
the projector onto the support of
$\sigma_k$, and the prim on the sum means
that the terms corresponding to $w_k=0$
are omitted.}

{\bf Proof.} Spectral forms $\forall k,
\enskip w_k>0:\enskip
\sigma_k=\sum_{l_k}s_{l_k}\ket{l_k}
\bra{l_k}\quad$ (all $s_{l_k}$ positive)
give a spectral form $\sigma =
\sum_k\sum_{l_k}w_ks_{l_k}\ket{l_k}\bra{l_k}$
of $\sigma$ on account of the
orthogonality assumed in (13) and the
zero convention. Since numerical
functions define the corresponding
operator functions via spectral forms,
one obtains further
$$log^e(\sigma
)\equiv
\sum_k\sum_{l_k}[log^e(w_ks_{l_k})]\ket{l_k}
\bra{l_k}=
\sum_k'\sum_{l_k}[log(w_k)+log(s_{l_k})]
\ket{l_k} \bra{l_k}=$$ $$
\sum_k'[log(w_k)]Q_k+\sum_k'
\sum_{l_k}[log(s_{l_k})]\ket{l_k}
\bra{l_k}.$$ (In the last step
$Q_k=\sum_{l_k}\ket{l_k}\bra{l_k}$ for
$w_k>0$ was made use of.) The same is
obtained from the RHS when the spectral
forms of $\sigma_k$ are substituted in
it. \hfill $\Box$

{\bf Proposition 1.} {\it Let condition
(12b) be valid for the states $\rho$ and
$\sigma$, and let an orthogonal state
decomposition (13) be given. Then one has
$$S(\rho||\sigma)=S\Big(\sum_kQ_k\rho
Q_k\Big)-S(\rho )+H(p_k||w_k)+\sum_kp_k
S(Q_k\rho
Q_k/p_k||\sigma_k),\eqno{(15)}$$ where,
for $w_k>0$, $Q_k$ projects onto the
support of $\sigma_k$, and $Q_k\equiv 0$
if $w_k=0$, $p_k\equiv \tr (\rho Q_k)$,
and
$$H(p_k||w_k)\equiv
\sum_k[p_klog(p_k)]-\sum_k[p_klog(w_k)]
\eqno{(16)}$$ is the classical discrete
counterpart of the quantum relative
entropy, valid because $(p_k>0)\enskip
\Rightarrow (w_k>0)$.}

One should note that the claimed validity
of the classical analogue of (12b) is due
to the definitions of $p_k$ and $Q_k$.
Besides, (13) implies that $(\sum_kQ_k)$
projects onto supp$(\sigma )$. Further,
as a consequence of (12b),
$(\sum_kQ_k)\rho =\rho$. Hence, $\tr
\Big(\sum_kQ_k\rho Q_k\Big)=\tr
(\sum_kQ_k\rho )=1$.

We call decomposition (15) {\it the
mixing property of relative entropy}.

{\bf Proof} of proposition 1: We define
$$\forall k,\enskip p_k>0:\quad
\rho_k\equiv Q_k\rho
Q_k/p_k.\eqno{(17)}$$ First we prove that
(12b) implies
$$\forall k,\enskip p_k>0:\quad
\mbox{supp}(\rho_k)\subseteq \mbox{supp}
(\sigma_k).\eqno{(18)}$$

Let $k$, $p_k>0$, be an arbitrary fixed
value. We take a pure-state decomposition
$$\rho
=\sum_n\lambda_n\ket{\psi_n}\bra{\psi_n}
\eqno{(19a)},$$  $\forall n:\enskip
\lambda_n>0$. Applying $Q_k...Q_k$ to
(19a), one obtains another pure-state
decomposition
$$Q_k\rho Q_k=p_k\rho_k
=\sum_n\lambda_nQ_k\ket{\psi_n}\bra{\psi_n}
Q_k\eqno{(19b)}$$ (cf (17)). Let
$Q_k\ket{\psi_n}$ be a nonzero vector
appearing in (19b). Since (19a) implies
that $\ket{\psi_n}\in \mbox{supp}(\rho )$
(cf Appendix 2(ii)), condition (12b)
further implies $\ket{\psi_n}\in
\mbox{supp}(\sigma )$. Let us write down
a pure-state decomposition
$$\sigma =\sum_m
\lambda'_m\ket{\phi_m}\bra{\phi_m}
\eqno{(20)}$$ with $\ket{\phi_1}\equiv
\ket{\psi_n}$. (This can be done with
$\lambda'_1>0$ cf \cite{Hadji}.) Then,
applying $Q_k...Q_k$ to (20) and taking
into account (13), we obtain the
pure-state decomposition
$$Q_k\sigma Q_k=w_k\sigma_k=\sum_m
\lambda'_mQ_k\ket{\phi_m}\bra{\phi_m}
Q_k.$$ (Note that $w_k>0$ because $p_k>0$
by assumption.) Thus,
$Q_k\ket{\psi_n}=Q_k\ket{\phi_1}\in
\mbox{supp}(\sigma_k)$. This is valid for
any nonzero vector appearing in (19b),
and these span supp$(\rho_k)$ (cf
Appendix 2(ii)). Therefore, (18) is
valid.

On account of (12b), the standard
logarithm can be replaced by the extended
one in definition (12a) of relative
entropy: $\quad S(\rho ||\sigma
)=-S(\rho)-\tr [\rho log^e(\sigma
)]\quad$. Substituting (13) on the RHS,
and utilizing (14), the relative entropy
$S(\rho ||\sigma )$ becomes
$$-S(\rho )-\tr \Big\{\rho
\Big[\sum_k'[log(w_k)]Q_k+\sum_k'[
log^e(\sigma_k)]\Big]\Big\}=-S(\rho
)-\sum_k'[p_klog(w_k)]-\sum_k'\tr [\rho
log^e(\sigma_k)].$$ Adding and
subtracting $H(p_k)$, replacing
$log^e(\sigma_k)$ by
$Q_k[log^e(\sigma_k)]Q_k$, and taking
into account (16) and (17), one further
obtains
$$S(\rho ||\sigma
)=-S(\rho )+H(p_k)+H(p_k||w_k)
-\sum_k'p_k\tr [\rho_klog^e(\sigma_k)].$$
(The zero convention is valid for the
last term because the density operator
$Q_k\rho Q_k/p_k$ may not be defined.
Note that replacing $\sum_k$ by $\sum_k'$
in (16) does not change the LHS because
only $p_k=0$ terms are omitted.)

Adding and subtracting the entropies
$S(\rho_k)$ in the sum, one further has
$$S(\rho ||\sigma
)=-S(\rho )+H(p_k)+H(p_k||w_k)+
\sum_k'p_kS(\rho_k)+\sum_k'p_k\{-S(\rho_k)
-\tr [\rho_klog^e(\sigma_k)]\}.$$
Utilizing the mixing property of entropy,
one can put $S\Big(\sum_kp_k\rho_k\Big)$
instead of
$[H(p_k)+\sum_k'p_kS(\rho_k)]$. Owing to
(18), we can replace $log^e$ by the
standard logarithm and thus obtain the
RHS(15). \hfill $\Box$

{\bf Remark.} {\it In a sense, (15) runs parallel
to Donald's identity
$$S(\rho||\sigma)=
\sum_kp_kS(\rho_k||\sigma )-H(p_k),$$ when an
orthogonal decomposition $\rho =\sum_kp_k\rho_k$
of the first state $\rho$ in relative entropy is
given.}

For a general decomposition $\rho
=\sum_kp_k\rho_k$ of the first state Donald's
identity reads
$$S(\rho ||\sigma
)=\sum_kp_kS(\rho_k||\sigma
)-\sum_kp_kS(\rho_k||\rho )$$ \cite{Donald},
\cite{Schum} (relation (5) in the latter). The
more special relation in the remark follows from
this on account of the relation that generalizes
the mixing property of entropy: If $\rho
=\sum_kp_k\rho_k$ is any state decomposition,
then $$S(\rho )= \sum_kp_k S(\rho_k||\rho
)+\sum_kp_kS(\rho_k)$$ is valid (cf Lemma 4 and
Remark 1 in \cite{Mutual}).

Now we turn to the derivation of some
consequences of proposition 1.

Let $\rho$ be a state and
$A=\sum_ia_iP_i+\sum_ja_jP_j$ a spectral
form of a discrete observable (Hermitian
operator) $A$, where the eigenvalues
$a_i$ and $a_j$ are all distinct. The
index $i$ enumerates all the detectable
eigenvalues, i. e., $\forall i:\enskip
\tr (\rho P_i)>0$, and $\tr [\rho
(\sum_iP_i)]=1$.

The simplest quantum measurement of $A$ in $\rho$
changes this state into the L\"{u}ders state:
$$\rho_L(A)\equiv \sum_iP_i\rho
P_i\eqno{(21)}$$ (cf (3a) and (3c)). Such a
measurement is often called "ideal".

{\bf Corollary 1.} {\it The relative-entropic
"distance" from any quantum state to its
L\"{u}ders state is the difference between the
corresponding quantum entropies:}
$$S\Big(\rho ||\sum_iP_i\rho
P_i\Big)=S\Big(\sum_iP_i\rho
P_i\Big)-S(\rho ).$$

{\bf Proof.} First we prove that
$$\mbox{supp}(\rho )\subseteq
\mbox{supp}\Big(\sum_iP_i\rho
P_i\Big).\eqno{(22)}$$ To this purpose,
we write down a decomposition (19a) of
$\rho$ into pure states. One has
$\mbox{supp}(\sum_iP_i)\supseteq
\mbox{supp}(\rho )$ (equivalent to the
certainty of $(\sum_iP_i)$ in $\rho$, cf
\cite{Roleof}), and the decomposition
(19a) implies that each $\ket{\psi_n}$
belongs to $\mbox{supp}(\rho )$ (cf
Appendix 2(ii)). Hence, $\ket{\psi_n}\in
\mbox{supp}(\sum_iP_i)$; equivalently,
$\ket{\psi_n}=(\sum_iP_i)\ket{\psi_n}$.
Therefore, one can write
$$\forall n:\quad \ket{\psi_n}=\sum_i(P_i
\ket{\psi_n}).\eqno{(23a)}$$ On the other
hand, (19a) implies
$$\sum_iP_i\rho
P_i=\sum_i\sum_n\lambda_nP_i\ket{\psi_n}
\bra{\psi_n}P_i.\eqno{(23b)}$$ As seen
from (23b), all vectors
$(P_i\ket{\psi_n})$ belong to
supp$(\sum_iP_i\rho P_i)$. Hence, so do
all $\ket{\psi_n}$ (due to (23a)). Since
$\rho$ is the mixture (19a) of the
$\ket{\psi_n}$, the latter span
$\mbox{supp}(\rho )$ (cf Appendix 2(ii)).
Thus, finally, also (22) follows.

In our case $\sigma \equiv \sum_iP_i\rho
P_i$ in (15). We replace $k$ by $i$.
Next, we establish
$$\forall i:\quad Q_i\rho Q_i=P_i\rho
P_i.\eqno{(24)}$$ Since $Q_i$ is, by
definition, the support projector of
$(P_i\rho P_i)$, and $P_i(P_i\rho
P_i)=(P_i\rho P_i)$, one has $P_iQ_i=Q_i$
(see Appendix 2(i)). One can write
$P_i\rho P_i=Q_i( P_i\rho P_i)Q_i$, from
which then (24) follows.

Realizing that $w_i\equiv \tr (Q_i\rho
Q_i)=\tr (P_i\rho P_i)\equiv p_i$ due to
(24), one obtains $H(p_i||w_i)=0$ and
$\quad \forall i:\quad S(Q_i\rho Q_i/p_i
||P_i\rho P_i/w_i)=0\quad$ in (15) for
the case at issue. This completes the
proof.\hfill $\Box$

Now we turn to a peculiar further
implication of Corollary 1.

Let $B=\sum_k\sum_{l_k}b_{kl_k}P_{kl_k}$
be a spectral form of a discrete
observable (Hermitian operator) $B$ such
that all eigenvalues $b_{kl_k}$ are
distinct. Besides, let $B$ be more
complete than $A$ or, synonymously, a
refinement of the latter. This, by
definition means that
$$\forall k:\quad
P_k=\sum_{l_k}P_{kl_k}\eqno{(25)}$$ is
valid. Here $k$ enumerates both the $i$
and the $j$ index values in the spectral
form of $A$.

Let $\rho_L(A)$ and $\rho_L(B)$ be the
L\"{u}ders states (21) of $\rho$ with
respect to $A$ and $B$ respectively.

{\bf Corollary 2.} {\it The states
$\rho$, $\rho_L(A)$, and $\rho_L(B)$ lie
on a straight line with respect to
relative entropy, i. e., $\quad
S\Big(\rho || \rho_L(B)\Big)=S\Big(\rho
||\rho_L(A)\Big)+S\Big(\rho_L(A))||
\rho_L(B)\Big)\quad$, or explicitly:}
$$S\Big(\rho
||\sum_i\sum_{l_i}(P_{il_i}\rho
P_{il_i})\Big)=S\Big(\rho
||\sum_i(P_i\rho P_i)\Big)+
S\Big(\sum_i(P_i\rho P_i)||
\sum_i\sum_{l_i}(P_{il_i} \rho
P_{il_i})\Big).$$

Note that all eigenvalues $b_{kl_k}$ of
$B$ with indices others than $il_i$ are
undetectable in $\rho$.

{\bf Proof.} Corollary 1 immediately
implies
$$S\Big(\rho ||\rho_L(B)\Big)
=\Big[S\Big(\rho_L(B)\Big)-
S\Big(\rho_L(A)\Big)\Big]+
\Big[S\Big(\rho_L(A)\Big)-S(\rho
)\Big],$$ and, as easily seen from (21),
$\rho_L(B)= \Big(\rho_L(A)\Big)_L(B)$ due
to $P_{il_i}P_{i'}=\delta_{i,i'}P_{il_i}$
(cf (25)).

\hfill $\Box$

\section{Properties of coherence
information}

To begin with, we notice in (10) that
$I_C$ depends on $\rho$ and $A$, actually
only on the eigenprojectors of the
latter.

As a consequence of (10), one can also
write the definition of $I_C$ in the form
of a relative entropy:
$$I_C=S\Big(\rho ||\sum_lP_l\rho
P_l\Big)\eqno{(26)}$$ as follows from
corollary 1.

It was proved long ago \cite{Lind} that
$S\Big(\sum_l P_l\rho P_l\Big)>S(\rho )$
if and only if $A$ and $\rho$ are
incompatible, and the two entropies are
equal otherwise. Thus, in case of
compatibility $[A,\rho ]=0$, $I_C$ is
zero, otherwise it is positive. This is
what we would intuitively expect.

It was proved in \cite{Roleof} (theorem 2
there) that
$$I_C=w_{inc}I_C\Big(\sum_l^{inc}a_lP_l,
(\sum_l^{inc}P_l)\rho (\sum_l^{inc}P_l)
/w_{inc}\Big),\eqno{(27)}$$ where "inc"
on the sum denotes summing only over all
those values of $l$ the corresponding
$P_l$ of which are incompatible with
$\rho$, and $\quad w_{inc}\equiv \tr
(\rho \sum_l^{inc}P_l)$.

This corresponds to an intuitive
expectation that the quantity $I_C$
should depend only on those
eigenprojectors $P_l$ of $A$ that do not
commute with $\rho$, and not at all on
those that do.

We obtain (27) as a special case of a much more
general result below (cf the theorem and
propositions 2 and 3).

We shall need another known concept. For the sake
of precision and clarity, we define it.

{\bf Definition 2.} {\it One says that a
discrete observable $\bar A=\sum_m\bar
a_m\bar P_m$ (spectral form in terms of
distinct eigenvalues $\bar a_m$) is
coarser than or a coarsening of
$A=\sum_la_lP_l$ if there is a
partitioning $\Pi$ in the set
$\{l:\forall l\}$ of all index values of
the latter
$$\Pi:\qquad
\{l:\forall l\}=\sum_mC_m,$$ such that
$$\forall m:\quad \bar P_m=\sum_{l\in
C_m}P_l$$ ($C_m$ are classes of values of
the index $l$, and the sum is the union
of the disjoint classes). One also says
that $A$ is finer than or a refinement of
$\bar A$.}

{\bf Theorem.} {\it Let $\bar A$ be any
coarsening of $A$ (cf definition 2). Then
$$I_C(A,\rho )=I_C(\bar A,\rho )+
\sum_m\Big[p_mI_C\Big(\bar P_mA,\bar P_m\rho \bar
P_m/p_m\Big)\Big],\eqno{(28)}$$ and $\forall
m:\enskip p_m\equiv \tr (\rho \bar P_m)$. (If
$p_m=0$, then, by the zero convention, the
corresponding $I_C$ in (28) need not be defined.
The product is by definition zero.)}

Before we prove the theorem, we apply
corollary 2 to our case.

Under the assumptions of the theorem, one
has $$S\Big(\rho ||\sum_l (P_l\rho
P_l)\Big)=S\Big(\rho ||\sum_m(\bar P_m
\rho \bar P_m)\Big)+ S\Big(\sum_m(\bar
P_m\rho \bar P_m)||\sum_l (P_l\rho
P_l)\Big).\eqno{(29)}$$

{\bf Proof} of the Theorem. On account of
(26), (29) takes the form
$$I_C(A,\rho )=I_C(\bar A,\rho )+I_C\Big(A,
\sum_m(\bar P_m\rho \bar P_m)\Big).\eqno{(30)}$$
Utilizing (10) for the second term on the RHS,
the latter becomes $S\Big(\sum_l(P_l\rho
P_l)\Big)-S\Big(\sum_m(\bar P_m\rho \bar
P_m)\Big)$. Making use of the mixing property of
entropy in both these terms, and cancelling out
$H(p_m)$ (cf (7b) {\it mutatis mutandis}), this
difference, further, becomes
$\sum_mp_mS\Big((\sum_{l\in C_m}P_l\rho
P_l)/p_m\Big)-\sum_mp_mS\Big(\bar P_m\rho \bar
P_m/p_m)\Big)$. Its substitution in (30) with the
help of (10) (and definition 2) then gives the
claimed relation (28). (Naturally, one must be
aware of the fact that $\bar A$ is a coarsening
of $A$, hence $\enskip \forall m:\enskip [\bar
P_m,A]=0,\enskip$ implying $\enskip A\equiv
\sum_m\sum_{m'}\bar P_mA\bar P_{m'}=\sum_m\bar
P_mA$.) \hfill $\Box$

If $\bar A$ is any coarsening of $A$, then the
index values $m$ of the former replace classes
$C_m$ of index values $l$ of the latter. Hence,
coherence in $\bar A$ - as a cooperative role of
index values - must be poorer than in $A$.
Therefore, one would intuitively expect that
$I_C(\bar A,\rho )$ must not be larger than
$I_C(A,\rho )$. The theorem confirms this, and
tells more: it gives the expression by which
$I_C(A,\rho )$ exceed $I_C(\bar A,\rho )$. One
wonders what the intuitive meaning of this is.

{\it Discussion of the theorem.} Let us think of
$\rho$ as describing a laboratory ensemble, and
let us imagine that an ideal measurement of $\bar
A$ is performed on each quantum system in the
ensemble. The ensemble $\rho$ is then replaced by
the mixture $\quad \sum_mp_m(\bar P_m\rho \bar
P_m/p_m)\quad$ of subensembles $\quad (\bar
P_m\rho \bar P_m/p_m)$. One can think of the
measurement of the more refined observable $A$ as
taking place in two steps: the first is the
mentioned measurement of the coarser observable
$\bar A$, and the second is a continuation of
measurement of $A$ in each subensemble $\quad
(\bar P_m\rho \bar P_m/p_m)$. Let us assume {\it
additivity} of $I_C$ in two-step measurement.

Further, let us bear in mind that, though $I_C$
is meant to be a property of each individual
member of the ensemble $\rho$, it is {\it
statistical}, i. e., it is given in terms of the
ensemble. Finally, in the second step we have an
ensemble of subensembles (a superensemble). Since
our system is anywhere in the entire ensemble
$\quad \sum_mp_m(\bar P_m\rho \bar P_m/p_m)\quad$
of the second step, one must average over the
superensemble with the statistical weights $p_m$
of its subensemble-members $\quad (\bar P_m\rho
\bar P_m/p_m)$.

If $m'\not= m$, then the part $\quad \bar
P_{m'}A\quad$ of $\quad A=\sum_{m''}\bar
P_{m''}A\quad$ is evidently undetectable in the
subensemble $\rho_m$. Hence, only $\bar P_mA$ is
relevant from the entire $A$, i. e., $I_C(A,\rho
)$ reduces to $I_C(\bar P_mA,\rho_m)$ there.

In this way one can understand relation (28).
What have we learnt from this? It is that $I_C$
is additive and statistical. This conclusion is
in keeping with the neighboring quantity
$S(A,\rho )$. Namely, one can easily derive a
relation similar to (28) for it:
$$S(A,\rho )=S(\bar A,\rho )+\sum_mp_mS(\bar
P_mA,\bar P_m\rho \bar P_m/p_m).$$ That $I_C$ and
$S(A,\rho )$ behave equally in an additive and
statistical way is no surprise since they are
terms in the same general decomposition (11) of
the entropy $S(\rho )$ of the state $\rho$.

The theorem is a substantially stronger form of a
previous result (theorem 3 in \cite{Roleof}), in
which $I_C(A,\rho )\geq I_C(\bar A,\rho )$ was
established with necessary and sufficient
conditions for equality, which are obvious in the
theorem. ($I_C$ was denoted by $E_C$ in previous
work, cf my comment following proposition 5
below.)

The theorem has the following immediate
consequences.

{\bf Proposition 2.} {\it If the coarsening $\bar
A$ defined in definition 2 is {\it compatible}
with $\rho$, then (28) reduces to}
$$I_C(A,\rho )= \sum_m\Big[p_mI_C\Big(\bar P_mA,\bar
P_m\rho \bar
P_m/p_m\Big)\Big].\eqno{(31)}$$

{\bf Proposition 3.} {\it Let us define a
coarsening $\Pi$ (cf definition 2) that
partitions $\{l:\forall l\}$ into at most
three classes: $C_{inc}$ comprising all
index values $l$ for which $a_l$ is
detectable (i. e., of positive
probability) and $P_l$ is incompatible
with $\rho$, $C_{comp}$ consisting of all
$l$ for which $a_l$ is detectable and
$P_l$ is compatible with $\rho$, and,
finally, $C_{und}$ which is made up of
all $l$ for which $a_l$ is undetectable.
The coarsening thus defined is compatible
with $\rho$, and (31) reduces to (27).}

{\bf Proof.} In the coarsening $\Pi$ of
proposition 3 the index $m$ takes on
three 'values': 'inc', 'comp', and 'und'.
It is easily seen that the coarser
observable $\bar A$ thus defined is
compatible with $\rho$. Hence, (31)
applies. Further, the second and third
terms are zero. In this way, (27) ensues.
\hfill $\Box$

{\bf Proposition 4.} {\it Coherence information
$I_C$ is unitary invariant, i. e., $\quad
I_C(A,\rho )=I_C(UAU^{\dagger},U\rho
U^{\dagger}),\quad$ where $U$ is an arbitrary
unitary operator.}

{\bf Proof.} Relative entropy is known to
be unitary invariant. On account of (26),
so is $I_C$.\hfill $\Box$

This is as it should be because $I_C$
should not depend on the basis in the
state space: $UAU^{-1}$ and $U\rho
U^{-1}$ can be understood as $A$ and
$\rho$ respectively viewed in another
basis.

{\bf Proposition 5.} {\it Coherence
information $I_C$ is convex.}

{\bf Proof.} This is an immediate
consequence of the known convexity of
relative entropy (cf (26)) under joint
mixing of the two states in it.

On account of convexity we know that $I_C$ is an
{\it information entity}, and not an entropy one
(or else it would be concave). In previous work
\cite{FHPR02}, \cite{Roleof}, \cite{ent-meas} the
same quantity (the RHS of (10)) was erroneously
denoted by $E_C(A,\rho )$ and treated as an
entropy quantity. But this does not imply that
any of the applications of $E_C(A,\rho )$ was
erroneous. All one has to do is to replace this
symbol by $I_C(A,\rho )$ and keep in mind that
one is dealing with an information quantity.

\section{Conclusion}

Perhaps it is of interest to comment upon the
more standard uses of the term "coherence" in the
literature.

One encounters the basic use of the word
"coherence" in the properties of light waves. One
distinguishes two types of coherence there: (i)
Temporal coherence, which is a measure of the
correlation between the phases of a light wave at
different points along the direction of
propagation, and (ii) spatial coherence, which is
a measure of the correlation between the phases
of a light wave at different points transverse to
the direction of propagation. (The fascinating
phenomenon of holography requires a large measure
of both temporal and spatial coherence of light.)

Quantum "coherence" refers also to large numbers
of particles that cooperate collectively in a
single quantum state. The best known examples are
superfluidity, superconductivity, and laser
light, all macroscopic phenomena. In the last
example different parts of the laser beam are
related to each other in phase, which can lead to
interference effects. "Coherence" is often
related to different kinds of correlations, see,
e. g., \cite{JS}.

In all mentioned examples "coherence" refers to
an {\it absolute} property of the quantum state
of the system; in contrast with the use of the
term in this article, which expresses a {\it
relative} property: relation between observable
and state. As it was mentioned, the kind of
quantum coherence studied in this article can be
more fully called "eigenvalue coherence of an
observable in relation to a state" in view of the
cooperative role of the eigenvalues (or rather
their quantum numbers, because the values of the
eigenvalues play no role) as seen in (4).

In the literature one often finds the claim that
quantum pure states are coherent. From the
analytical point of view of this article one can
say that a pure state $\ket{\psi }$ is {\it not
coherent} with respect to any observable for
which $\ket{\psi }\bra{\psi }$ is an
eigenprojector. But it is coherent with respect
to all other observables.

\subsection{On generality of the
results}

 A question may linger on to the
end of this study: What if the observable
is not a discrete one? Can one still
speak of eigenvalue coherence in relation
to a given state $\rho$?

It seems to me that the answer is that
one should write down the following
partial spectral form of a general
observable $A'$: $$A'=\sum_la_lP_l+
P^{\perp}A'P^{\perp},$$ where the
summation goes over {\it all eigenvalues}
of $A'$, and $P\equiv \sum_lP_l$. One
should take the discrete coarsening $A$
of $A'$: $$A\equiv
\sum_la_lP_l+aP^{\perp},$$ where the
eigenvalue $a$ is arbitrary but distinct
from all $\{a_l:\forall l\}$. Then the
expounded eigenvalue coherence theory
should by applied to $A$, and it should
be valid for $A'$ (as the best we can do
for the latter). In a preceding article
\cite{Roleof} the case when
$P^{\perp}\not= 0$ with the eigenvalue
$a$ undetectable was studied.

One has eigenvalue coherence of a general
observable $A'$ in relation to a state
$\rho$ if either $A'$ has at least two
eigenvalues or if $A'$ has at least one
eigenvalue and $P^{\perp}\not= 0$.

Another question that may linger on is
whether the state $\rho$ that was used in
this paper is really general. If $\rho$
has an infinite-dimensional range and $A$
has infinitely-many eigenvalues, it may
happen that there are infinitely-many
detectable ones. The expounded theory
covers also this case.

\subsection{Summing up}

In an attempt to understand the essential
features of two-slit interference (see lemma 1
followed by its application to two-slit
interference in subsection 1.2), a general
coherence theory was developed based on the
assumption that 'coherence' equals
'incompatibility' $[A,\rho ]\not= 0$ between
observable and state. Since this relation means
that $\rho$ is incompatible with at least one
eigenevent (eigenprojector) $P_l$ of $A$, and
this property is independent of the eigenvalues,
it was argued that the entire family of
observables with one and the same decomposition
of the identity $\sum_lP_l=I$ (the latter is
called "closure relation" if $A$ is complete)
should have the same amount of incompatibility.
This discarded the Wigner-Yanase-Dyson family of
skew informations (6). Further, it was argued
that the necessarily nonnegative quantity
$\enskip S(A^c,\rho )-S(\rho )\enskip$ was a
natural measure of incompatibility between a
complete observable $A^c$ and the state $\rho$
satisfying the stated claim. Finally,
interpolating between the case of a complete and
that of a compatible observable (see (8), (9) and
(10)), the general expression (10) was obtained.

Thus, a natural quantum measure of how much of
coherence, and, equivalently, incompatibility,
there is if a discrete observable
$A=\sum_la_lP_l$ and a state $\rho$ are given was
derived along the expounded argument. It was
called coherence or incompatibility information
(denoted by $I_C(A,\rho )$ or shortly $I_C$) in
section 2.

A deviation into a general
relative-entropy investigation was made
in section 3. What was called 'the mixing
property of relative entropy'
(parallelling that of entropy) was
derived, and so were two corollaries.

The relative-entropy results were utilized to
express coherence information $I_C(A,\rho )$ in
the form of a relative entropy (cf (26)) in
section 4. Connection between the coherence
information $I_C(\bar A,\rho )$ of any coarsening
$\bar A$ (cf definition 2) of an observable $A$
and $I_C(A,\rho )$ was obtained in the theorem.
Its intuitive meaning was discussed. It was
concluded that $I_C$ is additive in two-step
measurement and statistical.

The corresponding relation took a much simpler
form in case $\bar A$ was compatible with $\rho$
(cf proposition 2). In a special case of this a
result from previous work was recognized (cf
proposition 3 and (27)). Coherence information
was shown to be unitary invariant (proposition 4)
and convex (proposition 5).

In previous work \cite{FHPR02}, \cite{Roleof},
\cite{ent-meas} the coherence information $I_C$
was successfully utilized in analyzing bipartite
quantum correlations. The last one of them filled
in an information-theoretical gap noted in
preceding investigation of the measurement
process \cite{Vedral}.

Since a number of new properties of $I_C$ have
now been obtained, even more fruitful
applications can be expected.\\

\noindent {\bf Appendix 1.}\\

\indent We prove the equivalence of the
negations of the four claims in lemma 1.
("$\neg$ (i)" is the negation of (i)
etc., and "$(\Leftrightarrow )$" is the
claim of "$\Leftrightarrow$") The logical
scheme of the proof is: $\neg$ (ii)
$\Leftrightarrow$ $\neg$ (iii)
$\Leftrightarrow$ $\neg$ (iv); $\neg$
(ii) $\Rightarrow$ $\neg$ (i)
$\Rightarrow$ $\neg$ (iii).

$\neg$ (ii) $(\Leftrightarrow )$ $\neg$
(iii): One can always write $\rho
=\sum_l\sum_{l'}P_l\rho P_{l'}$. Since
$A$ and $\rho$ commute if and only if
each eigenprojector $P_l$ of $A$ commutes
with $\rho$, the claimed equivalence is
obvious.

$\Big(\neg$ (iii) $\Rightarrow$ $\neg$
(iv)\Big) is obvious. To prove
\Big($\neg$ (iv) $\Rightarrow$ $\neg$
(iii)\Big), we restrict the operators $B$
to ray projectors $\ket{a}\bra{a}$. Then
$\neg$ (iv) implies $\tr (\rho
\ket{a}\bra{a})=\bra{a}\rho
\ket{a}=\bra{a}\rho_L\ket{a}$ for every
state vector $\ket{a}$. But then, as well
known, one must have $\rho =\rho_L$,
which is $\neg$ (iii).

$\neg$ (ii) $(\Rightarrow )$ $\neg$ (i):
In view of $\rho =\sum_l\sum_{l'}P_l\rho
P_{l'}$, commutation of $\rho$ with each
$P_l$ implies $\neg$ (i).

$\neg$ (i) $(\Rightarrow )$ $\neg$ (iii):
Let us assume that $\rho
=\sum_lp_l\rho_l$, and that each state
$\rho_l$ has the sharp value of the
corresponding eigenvalue $a_l$ of $A$.
Then $\rho_l =P_l\rho_lP_l$ (cf lemma
A.4. in \cite{FHFoundPL}). Substituting
this in the state decomposition, and
subsequently evaluating $\rho_L$
according to (3a)-(3c), one can see that
$\neg$ (iii) follows.\hfill
$\Box$\\

\noindent {\bf Appendix 2.}\\

Let $\rho =\sum_n\lambda_n\ket{n}\bra{n}$
be an arbitrary decomposition of a
density operator into ray projectors, and
let $E$ be any projector. Then $$E\rho
=\rho \quad \Leftrightarrow \quad \forall
n:\enskip E\ket{n}=\ket{n}\eqno{(A.1)}$$
(cf Lemma A.1. and A.2. in
\cite{FHJP94}).

(i) If the above decomposition is an
eigendecomposition with positive weights,
then $\sum_n\ket{n}\bra{n}=Q$, $Q$ being
now the support projector of $\rho$, and,
on account of (A.1),
$$E\rho =\rho \quad \Rightarrow \quad
EQ=Q.\eqno{(A.2)}$$.

(ii) Since one can always write $Q\rho
=\rho$, (A.1) implies that all $\ket{n}$
in the arbitrary decomposition belong to
supp$(\rho )$. Further, defining a
projector $F$ so that supp$(F)\equiv$
span$(\{\ket{n}:\forall n\})$, one has
$FQ=F$. Equivalence (A.1) implies $F\rho
=\rho$. Hence, (A.2) gives $QF=Q$.
Altogether, $F=Q$, i. e., the unit
vectors $\{\ket{n}:\forall n\}$
span supp$(\rho)$.\\

\noindent {\bf References}\\

\end{document}